\documentclass[12pt]{article}

\usepackage{graphicx}
\usepackage{amsmath}
\usepackage{amssymb}

\usepackage{color}
\input{colordvi.tex}

\def\be{\begin{eqnarray}}
\def\ee{\end{eqnarray}}
\def\ne{\nonumber\end{eqnarray}}
\def\nn{\nonumber}
\def\bra#1{\left<{#1}\right|}
\def\ket#1{\left|{#1}\right>}
\def\vac{\left|{0}\right>}
\def\vev#1#2{\left<{#1}|{#2}\right>}

\def\a{\alpha}

\newcommand{\gsim}{\mathrel{\mathop{\kern-0pt >}\limits_{\sim}}}
\newcommand{\lsim}{\mathrel{\mathop{\kern-3pt <}\limits_{\sim}}}
\def\sitarel#1#2{\mathrel{\mathop{\kern-0pt #1}\limits_{#2}}}

\begin{document}

\vspace*{-1cm}\hspace*{12.5cm}
\hfill\parbox{4cm}
{\normalsize 
{RIKEN-TH-128}\\
{UT-08-12}\\
{April, 2008}
}\\

\vskip .3in
\centerline{\Large\bf 
              Schnabl's Solution and Boundary States 
}
\vskip .1in
\centerline{\Large\bf 
              in Open String Field Theory
}

\vskip .4in
\centerline{\sc Teruhiko Kawano$^{\star}$, 
                Isao Kishimoto$^{\dag}$ 
                and 
                Tomohiko Takahashi$^{*}$}

\vskip .2in
\centerline{\it ${}^{\star}$Department of Physics, University of Tokyo, Hongo, 
Tokyo 113-0033, Japan}
\vskip 0in
\centerline{\it ${}^{\dag}$Theoretical Physics Laboratory, RIKEN, 
Wako 351-0198, Japan}
\vskip 0in
\centerline{\it ${}^{*}$Department of Physics, Nara Women's University,
Nara 630-8506, Japan}

\vskip .4in
\centerline{\small \bf Abstract}
\medskip
\noindent{\small 
We discuss that Schnabl's solution is an off-shell extension of 
the boundary state describing a D-brane in the closed string sector. 
It gives the physical meaning of the gauge invariant overlaps for the solution 
in our previous paper and supports Ellwood's recent proposal in the operator 
formalism. 
}

\vskip .5in

In our previous paper \cite{KKT}, it was discussed that 
a class of gauge invariant observables gives the same values 
for the analytic solution \cite{Schnabl} given by Schnabl as the ones 
for the numerical solution \cite{SZ,MT,GR} in the level truncation 
in open string field theory \cite{Witten}. 
It gives another interesting evidence that 
the numerical solution is gauge equivalent to the analytic solution. 

The gauge invariant observables ${\cal O}_V\left(\Psi\right)$ for 
an open string field $\Psi$ are called 
gauge invariant overlaps in \cite{KKT} and are given by 
\be
{\cal O}_V(\Psi)&=&\left< V(i)f_{\cal I}[\Psi]\right>
=\bra{{\cal I}}V(i)\ket{\Psi},
\qquad
f_{\cal I}(z)\equiv \frac{2z}{1-z^2},
\label{overlap}
\ne
where the CFT correlator is defined on an upper half plane. 
The on-shell closed string vertex operator $V(i)$ is inserted at the 
midpoint of the open string, and the conformal mapping $f_{\cal I}(z)$ 
plays the role of the identity state $\bra{{\cal I}}$, identifying 
the left half of the string with its right half. 
They were originally discussed in \cite{Zwiebach} to give the 
interaction of an on-shell closed string with open strings in the 
open string field theory. 

For the analytic solution
\be
\ket{\Psi_{\lambda=1}}
=\ket{\psi_0}
+\sum_{n=0}^{\infty}\left(\ket{\psi_{n+1}}-\ket{\psi_n}
-\partial_r\ket{\psi_r}|_{r=n}\right)
\equiv\ket{\psi_0}+\ket{\chi},
\ne
since it was shown \cite{KKT,Ellwood} that 
${\cal O}_V(\psi_r)$ doesn't depend on $r$, one can see that 
\be
{\cal O}_V(\Psi_{\lambda=1})
=\bra{{\cal I}}V(i)\ket{\Psi_{\lambda=1}}
=\bra{{\cal I}}V(i)\ket{\psi_0}.
\ne
Therefore, it suggests that only the first term $\ket{\psi_0}$ contributes 
to the gauge invariant overlaps ${\cal O}_V(\Psi_{\lambda=1})$.

In the paper \cite{KKT}, the gauge invariant overlaps were used to 
examine whether the numerical solution is gauge equivalent to the analytic 
solution, but their physical meaning wasn't clear. 
Recently, Ellwood has made an interesting proposal \cite{Ellwood} that  
the gauge invariant overlaps are related to the closed string tadpoles as
\be
{\cal O}_V(\Psi)={\cal A}_{\Psi}(V)-{\cal A}_{0}(V),
\ne
where ${\cal A}_{\Psi}(V)$ is the disk amplitude for a closed string 
vertex operator $V$ with the boundary condition of the CFT given 
by the open string field solution $\Psi$, and ${\cal A}_{0}(V)$ is 
the usual disk amplitude in the perturbative vacuum. 

In fact, for the analytic solution, he has shown that this is the case. 
Since, after tachyon condensation, there are no D-branes, no closed 
tadpoles are available, and thus ${\cal A}_{\Psi}(V)=0$. Therefore, 
the gauge invariant overlap ${\cal O}_V(\Psi)$ for the solution gives 
the usual disk amplitude of the opposite sign with one closed string emitted. 
It suggests that the analytic solution given by Schnabl is closely related to 
the boundary state describing a D-brane in the closed string perturbation 
theory, as expected. 

In this note, we will explicitly demonstrate this in the operator formalism of 
open string field theory by using the Shapiro-Thorn vertex 
$\langle\hat\gamma(1_{\rm c},2)|$ \cite{S-T}, which maps an open string
state to 
the corresponding closed string state. (For more detail, see appendix B in 
\cite{KKT}.) 
By using the open-closed string vertex $\langle\hat\gamma(1_{\rm c},2)|$, 
as discussed in detail in \cite{KKT}, 
one may rewrite a gauge invariant overlap as
\be
{\cal O}_V(\psi)
=\left<\hat\gamma(1_{\rm c},2)|\phi_{\rm c}\right>_{1_{\rm c}}|\psi\rangle_2,
\ne
where $|\phi_{\rm c}\rangle_{1_{\rm c}}$ is an on-shell state given by 
the vertex operator $V$ of the closed 
string $1_c$, and $|\psi\rangle_2$ is a state of the open string $2$.

For the analytic solution, one has
\be
{\cal O}_V(\Psi_{\lambda=1})
&=&\left<\hat\gamma(1_{\rm c},2)|
\phi_{\rm c}\right>_{1_{\rm c}}|\Psi_{\lambda=1}\rangle_2
\nn\\
&=&
\left<\hat\gamma(1_{\rm c},2)|
\phi_{\rm c}\right>_{1_{\rm c}}|\psi_0\rangle_2
+\left<\hat\gamma(1_{\rm c},2)|
\phi_{\rm c}\right>_{1_{\rm c}}|\chi\rangle_2,
\ne
and as mentioned above, the second term on the right hand side is zero 
for the on-shell closed string state $|\phi_{\rm c}\rangle_{1_{\rm c}}$. 
As for the first term on the right hand side, one can see that 
the open string tachyon state $\ket{\psi_0}=(2/\pi)c_1\vac$ is transformed 
via the vertex $\langle\hat\gamma(1_{\rm c},2)|$ 
into the boundary state $\bra{B\,}$. In fact, one can obtain the relation 
\be
\vev{\hat{\gamma}(1_c,2)}{\psi_0}_2{\cal P}_{1_c}
={1\over2\pi}\bra{B\,}c_{0}^{-}
\ne
with the level matching projection ${\cal P}_{1_c}$ for the closed string $1_c$, 
where the boundary state is the usual one
\be
\bra{B\,}=\bra{0}c_{-1}\bar{c}_{-1}c_0^{+}
e^{-\sum_{n=1}^{\infty}
[{1\over{n}}\a_n\cdot\bar\a_n+c_n\bar{b}_n+\bar{c}_n{b}_n]}
\ne
in the closed string perturbation theory. 
One thus finds that 
\be
{\cal O}_V(\Psi_{\lambda=1})= {1\over2\pi}\bra{B\,}c_{0}^{-}
\ket{\phi_{\rm c}},
\ne
which is in precise agreement with Ellwood's result. 

Since the second term $\left<\hat\gamma(1_{\rm c},2)|
\phi_{\rm c}\right>_{1_{\rm c}}|\chi\rangle_2$ doesn't necessarily vanish 
for off-shell closed string states, one may conclude that 
the transform of the analytic solution $\ket{\Psi_{\lambda=1}}$ 
via the Shapiro-Thorn vertex is an off-shell extension of the boundary state 
$\ket{B}$. Although it seems more elaborate to calculate 
$\left<\hat\gamma(1_{\rm c},2)|\chi\right>_2$, it would be interesting 
to find the relation of the off-shell boundary state with the equation of motion 
in closed string field theory \cite{Saadi,NonPoly,CSFT}.

Furthermore, given all the interactions between open strings and 
closed strings in the open-closed string field theory \cite{OpenClosed}, 
one may raise a question whether the Schnabl solution is 
consistent with the equations of motion of the open-closed string field 
theory, even in the vanishing string coupling constant limit, 
but it is beyond the scope of this paper. However, if it is consistent, 
the relation of the transform of the analytic solution 
$\ket{\Psi_{\lambda=1}}$ via the Shapiro-Thorn vertex with the boundary 
state $\ket{B}$ could be clearer along with the interactions of 
the theory \cite{OpenClosed}. We think that our observation in this paper 
may serve as an encouraging step in the investigation.\footnote{
More recently, the idea of this paper is extended to the marginal 
solutions \cite{Marginal} and the rolling tachyon solution \cite{Tachyon}
by one of the authors \cite{Kishimoto}.
It may suggest that our observation of this paper could be more
generic. 
}


\vskip .4in
\centerline{\bf Acknowledgement}
\bigskip
We are grateful to Taichiro Kugo for relevantly useful discussions. 
The work of T.~K. was supported in part by a Grant-in-Aid (\#19540268) 
from the MEXT of Japan.
The work of I.~K. was supported in part by the Special Postdoctoral 
Researchers Program at RIKEN and in part by a Grant-in-Aid (\#19740155) 
from the MEXT of Japan.
The work of T.~T. was supported in part by a Grant-in-Aid (\#18740152) 
from the MEXT of Japan.

\vskip .4in

\break


\begin{thebibliography}{99}

\bibitem{KKT}
  T.~Kawano, I.~Kishimoto and T.~Takahashi,
  ``Gauge Invariant Overlaps for Classical Solutions in Open String Field
  Theory,''
  Nucl.\ Phys.\  B {\bf 803}, 135 (2008),
  {\tt arXiv:0804.1541}.

\bibitem{Schnabl}
  M.~Schnabl,
``Analytic Solution for Tachyon Condensation in Open String Field Theory,''
  Adv.\ Theor.\ Math.\ Phys.\  {\bf 10}, 433 (2006), 
  {\tt hep-th/0511286}.

\bibitem{SZ}
  A.~Sen and B.~Zwiebach,
  ``Tachyon Condensation in String Field Theory,''
  JHEP {\bf 0003}, 002 (2000), 
  {\tt hep-th/9912249}.

\bibitem{MT}
  N.~Moeller and W.~Taylor,
  ``Level Truncation and the Tachyon in Open Bosonic String Field Theory,''
  Nucl.\ Phys.\  B {\bf 583}, 105 (2000), 
  {\tt hep-th/0002237}.

\bibitem{GR}
  D.~Gaiotto and L.~Rastelli,
``Experimental String Field Theory,''
  JHEP {\bf 0308}, 048 (2003), 
  {\tt hep-th/0211012}.

\bibitem{Witten}
  E.~Witten,
  ``Noncommutative Geometry and String Field Theory,''
  Nucl.\ Phys.\  B {\bf 268}, 253 (1986).

\bibitem{Zwiebach}
  B.~Zwiebach,
``Interpolating String Field Theories,''
  Mod.\ Phys.\ Lett.\  A {\bf 7}, 1079 (1992), 
  {\tt hep-th/9202015}.

\bibitem{Ellwood}
  I.~Ellwood,
  ``The Closed String Tadpole in Open String Field Theory,''
  JHEP {\bf 0808}, 063 (2008),
  {\tt arXiv:0804.1131}.

\bibitem{S-T}
  J.~A.~Shapiro and C.~B.~Thorn,
``Closed String - Open String Transitions And Witten's String Field Theory,''
  Phys.\ Lett.\  B {\bf 194}, 43 (1987).

\bibitem{Saadi}
  M.~Saadi and B.~Zwiebach,
  ``Closed String Field Theory from Polyhedra,''
  Annals Phys.\  {\bf 192}, 213 (1989).

\bibitem{NonPoly}
  T.~Kugo, H.~Kunitomo and K.~Suehiro,
  ``Nonpolynomial Closed String Field Theory,''
  Phys.\ Lett.\  B {\bf 226}, 48 (1989);\\
  T.~Kugo and K.~Suehiro,
  ``Nonpolynomial Closed String Field Theory: Action and its Gauge
  Invariance,''
  Nucl.\ Phys.\  B {\bf 337}, 434 (1990).

\bibitem{CSFT}
  B.~Zwiebach,
  ``Closed String Field Theory: Quantum Action and the B-V Master Equation,''
  Nucl.\ Phys.\  B {\bf 390}, 33 (1993), 
  {\tt hep-th/9206084}.

\bibitem{OpenClosed}
  B.~Zwiebach,
  ``Oriented Open-Closed String Theory Revisited,''
  Annals Phys.\  {\bf 267}, 193 (1998), 
  {\tt arXiv:hep-th/9705241}.

\bibitem{Marginal}
  M.~Schnabl,
  ``Comments on marginal deformations in open string field theory,''
  Phys.\ Lett.\  B {\bf 654}, 194 (2007)
  [arXiv:hep-th/0701248];\\
  M.~Kiermaier, Y.~Okawa, L.~Rastelli and B.~Zwiebach,
  ``Analytic solutions for marginal deformations in open string field theory,''
  JHEP {\bf 0801}, 028 (2008)
  [arXiv:hep-th/0701249].

\bibitem{Tachyon}
  S.~Hellerman and M.~Schnabl,
  ``Light-like tachyon condensation in Open String Field Theory,''
  {\tt arXiv:0803.1184}.

\bibitem{Kishimoto}
  I.~Kishimoto,
  ``Comments on Gauge Invariant Overlaps for Marginal Solutions in 
    Open String Field Theory,''
  {\tt arXiv:0808.0355}.

\end{thebibliography}
\end{document}